# Fiber optic computing using distributed feedback

Brandon. Redding[1], Joseph B. Murray[1], Joseph D. Hart[1], Zheyuan Zhu[2], Shuo S. Pang[2], Raktim Sarma[3]

[1]US Naval Research Laboratory, Washington, DC, USA
[2]CREOL, The College of Optics and Photonics, University of Central Florida, Orlando, FL, USA
[3]Center for Integrated Nanotechnologies, Sandia National Laboratories, Albuquerque, NM, USA

**Abstract**

The widespread adoption of machine learning and other matrix intensive computing algorithms has inspired renewed interest in analog optical computing, which has the potential to perform large-scale matrix multiplications with superior energy scaling and lower latency than digital electronics. However, most existing optical techniques rely on spatial multiplexing to encode and process data in parallel, requiring a large number of high-speed modulators and detectors. More importantly, most of these architectures are restricted to performing a single kernel convolution operation per layer. Here, we introduce a fiber-optic computing architecture based on temporal multiplexing and distributed feedback that performs multiple convolutions on the input data in a single layer (i.e. grouped convolutions). Our approach relies on temporally encoding the input data as an optical pulse train and injecting it into an optical fiber where partial reflectors create a series of delayed copies of the input vector. In this work, we used Rayleigh backscattering in standard single mode fiber as the partial reflectors to encode a series of random kernel transforms. We show that this technique effectively performs a random non-linear projection of the input data into a higher dimensional space which can facilitate a variety of computing tasks, including non-linear principal component analysis, support vector machines, or extreme learning machines. By using a passive fiber to perform the kernel transforms, this approach enables efficient energy scaling with orders of magnitude lower power consumption than GPUs, while using a high-speed modulator and detector maintains low latency and high data-throughput. Finally, our approach is readily integrated with fiber-optic communication links, enabling additional applications such as processing remote sensing data transmitted in the analog domain.

**Introduction**

Neural network-based machine learning algorithms have advanced dramatically in the last decade and are now routinely applied to a wide range of applications. However, this increase in performance has been accompanied by rapidly increasing computing demands—particularly in terms of the energy required to train and run these algorithms [1]. This has inspired research in alternative platforms capable of performing the computationally-intensive matrix-vector multiplications (MVMs) and kernel convolution operations at the heart of most machine learning algorithms more efficiently. Among these approaches, optical computing is particularly promising due to its superior energy scaling and potential to overcome memory-access bottlenecks [2–6].

These unique features have led to a series of impressive demonstrations in which photonic computing systems have performed benchmark tasks with comparable accuracy to digital electronic neural networks while consuming orders of magnitude less energy [3,7,8]. Most of these photonic computing schemes rely on spatial multiplexing in which input data is encoded in parallel on an array of modulators and the MVM output is recorded on an array of photodetectors. This approach has been explored both in free-space and integrated photonic platforms. Free-space platforms typically employ spatial light modulators (SLMs) to encode data and cameras to record the computed output. While this enables very large-scale computing (e.g. input vectors as large as $N \sim 10^6$ [9]), the latency is limited by the SLM and camera speeds. Integrated photonic solutions are both more compact and have the potential for higher-speed by exploiting state-of-the-art modulators and detectors [10,11]. However, processing large matrices on-chip remains a challenge due to the size, heat, and complexity of integrating large numbers of individually addressable modulators. These limitations have inspired recent proposals for temporally multiplexed architectures in which a single modulator is used to encode an entire vector [12–14]. These schemes could fill a gap in the photonic computing design space between the relatively slow, but large-scale free-space computing platforms and the high-speed but smaller-scale integrated photonic approaches. However, these temporally multiplexed architectures were either limited to a single neuron [12,15] or rely on complex optical routing schemes and have yet to be implemented experimentally [13,14].

In addition, most of these photonic approaches are unable to natively perform grouped convolutions in which multiple kernel operations are applied to the same input data. Grouped convolutions have been used in a variety of machine learning techniques (e.g. convolutional neural networks) due to their ability to extract hierarchical features in a dataset [16,17]. To perform grouped convolutions, existing photonic computing platforms would need to generate multiple copies of the input data, separately apply a kernel transform to each copy, and then recombine the outputs in the next layer—requiring a complex combination of beam splitters and routing optics with limited scalability. Recently, a time-wavelength multiplexing scheme was proposed to address this challenge [18]. This technique performed grouped convolutions in parallel using optical frequency combs by encoding distinct kernel transforms on different sets of comb teeth. After wavelength de-multiplexing, an array of photodetectors was used to record the output of each kernel transform, enabling impressive throughput at the cost of increased system complexity and limited scalability (the number of kernels operations was limited by the number of individually addressable comb teeth).

In this work, we introduce a temporally multiplexed optical computing platform that performs grouped convolutions using a simple and scalable approach based on distributed feedback in single mode fiber. We first encode the input vectors in the time domain as a pulse-train using a single high-speed modulator. This pulse-train is then injected into an optical fiber where a series of partial reflectors provide distributed feedback, generating a series of delayed copies of the input vector each weighted by the strength of a different reflector. In this demonstration, we rely on Rayleigh backscattering in standard single-mode fiber to provide this distributed feedback. Each Rayleigh scattering center creates a delayed copy of the input vector with random amplitude and phase—corresponding to the weights of a transformation matrix (i.e. a random kernel). The backscattered light is then recorded on a single, high-speed photodetector, performing the accumulation operation and introducing a non-linear transform. As explained in detail below, if the fiber is longer than the equivalent length of the encoded pulse train, then the fiber can perform multiple, distinct kernel operations on the input vector without requiring any additional routing or re-encoding of the input data. In principle, the weights of the transformation matrices (i.e. the kernels) could be inverse designed for specific computational tasks. As proof-of-concept, in this work, we use multiple random transformations to compute a non-linear random projection of the input vector into a higher dimensional space. We show that applying multiple random projections on the same input data can accelerate a variety of computing tasks including both unsupervised learning tasks such as non-linear principal component analysis (PCA) and supervised tasks including support vector machines (SVM) and extreme learning machines (ELM). More generally, this approach offers 5 major benefits: (1) It natively performs grouped convolutions. (2) It is scalable and is capable of processing relatively large-scale matrices (we demonstrate matrix operations on vectors with 784 elements) while maintaining high-speed (10 $\mu s$ per MVM). (3) Since this approach relies on a passive transform to perform the matrix operations, the energy consumption scales as $O(N)$, enabling significant reductions in power consumption compared to a graphics processing unit (GPU). (4) The entire system is constructed using commercially available, fiber-coupled components, enabling a robust and compact platform. (5) Since the system operates directly on fiber-coupled, time-series data, this approach could be used to directly analyze data transmitted over fiber, opening up additional applications in remote sensing, RF photonics, and telecommunications.

**Operating Principle**

The basic operating principle is outlined in Fig. 1. Data is first encoded in the time domain as a series of optical pulses. This pulse train is then injected into an optical fiber where it is partially reflected by a series of Rayleigh scattering centers. This distributed backscattering process randomly mixes the elements in the input vector, resulting in a backscattered speckle pattern that contains a series of random projections of the input vector. The Rayleigh backscattered (RBS) light is then recorded on a high-speed photodetector which performs a non-linear transform on the backscattered electric field [19,20]. The digitized speckle pattern can then be used for a variety of computing tasks including non-linear principal component analysis, support vector machines, or extreme learning machines. More broadly, the Rayleigh backscattering process can form the first layer of an artificial neural network, efficiently expanding the input data to higher dimensional space. Since the specific weights and connections in the first layers of a neural network are not critical in most applications [21], this process can be used to accelerate one of the most computationally intensive tasks in a neural network. A digital electronics back-end can then be used to complete the neural network. In recent years, random projections have been proposed for a variety of computing tasks in both free-space [9,19,20,22,23] and integrated photonic platforms [24,25]. While counterintuitive, researchers have shown that random transforms maintain key features in a dataset such as orthogonality while facilitating data analysis tasks such as dimensionality reduction or compressive sensing [26,27].

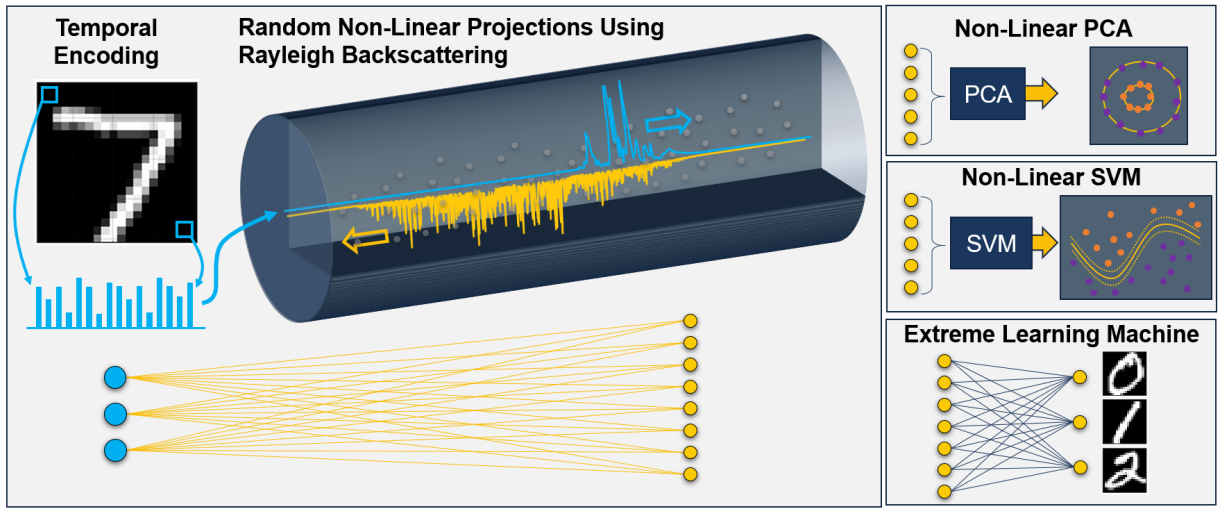

**Figure 1. Operating Principle.** Data is encoded in in the time-domain as a train of optical pulses. This pulse train is then injected into an optical fiber where distributed feedback is mediated by Rayleigh backscattering. The distributed feedback produces a series of delayed copies of the original data with random phase and amplitude. The backscattered signal is recorded on a photodetector, resulting in a random non-linear projection of the original data into a higher dimensional space, facilitating a variety of computing and data analysis tasks including non-linear PCA, SVM, or ELM.

Experimentally, this basic approach was realized using the architecture shown in Fig. 2(a). A continuous wave (CW) laser was coupled into an electro-optic modulator (EOM) which was used to encode the input data. An $N$-element vector, $A_N$, was encoded in the amplitude of a train of $N$ pulses, as shown in the inset of Fig. 2(a). This pulse train was coupled through a circulator into standard single-mode optical fiber where it was partially reflected by a series of Rayleigh scattering centers, creating a series of time-delayed copies of the original vector with random (though fixed) weights. The backscattered field, represented by an $M$-element complex vector $\tilde{C}_M$, was directed to a photodetector which performed a non-linear transform, generating photocurrent proportional to $|\tilde{C}_M|^2$. Figure 2(d,e) show examples of an encoded pulse train and the resulting Rayleigh backscattered speckle pattern, illustrating the dramatic increase in dimensionality provided by the Rayleigh backscattering process. In this case, the input data was a 60-element vector representing a SONAR signal (discussed in the SVM section below) and the backscattered pattern was a ~2000-element vector.

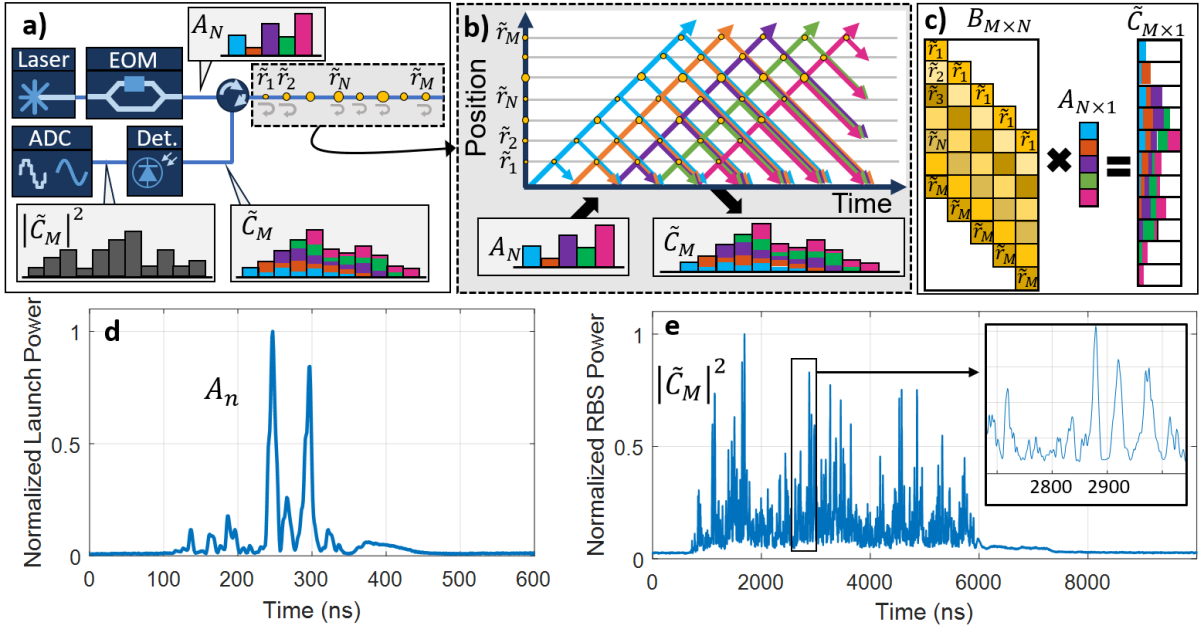

**Figure 2. Experimental Architecture.** (a) Data is encoded in the time domain using an EOM as a train of optical pulses which are injected into the fiber through a circulator. Rayleigh scattering provides distributed feedback and the backscattered field is then recorded on a photodetector, providing a non-linear response. (b) The distributed scattering process can be described using a space-time diagram which tracks the position of each pulse as it travels through the fiber. The pulses (shown in different colors for clarity) are partially reflected by Rayleigh scattering centers with varying complex reflectivity as they propagate down the fiber. As a result, the backscattered field contains randomly weighted contributions from each input pulse (i.e. each element in the input vector). (c) The distributed scattering process can be expressed as the multiplication between a complex transfer matrix $B$ and the input vector $A$. (d) An example pulse train representing the SONAR data discussed in the SVM section and (e) the resulting RBS pattern.

The random kernel transform introduced by the Rayleigh backscattering process can be visualized using the space-time diagram shown in Fig. 2(b). This diagram tracks the position of each pulse over time as it travels to the end of the fiber along a diagonal path in the upper-right direction. As each pulse propagates, it is partially reflected by a series of Rayleigh scattering centers and the reflected light travels back to the beginning of the fiber (along a diagonal path toward the bottom-right of the diagram). The Rayleigh backscattering at a given position in the fiber can be described by a complex reflectance, $\tilde{r}_k$, which is random but fixed. As shown in Fig. 2(b), the backscattered light at time $m$ includes contributions from each input pulse (once all of the pulses have entered the fiber) weighted by the complex reflectance at different positions in the fiber. While Fig. 2(b) presents a simplified description of the Rayleigh backscattering process as a series of discrete partial reflectors, in reality, Rayleigh scattering is effectively continuous. However, the temporal correlation width of the Rayleigh backscattered light is set by the bandwidth (or pulse duration) of the encoded data [28]. Thus, for the 5 ns pulses used in this work, the Rayleigh backscattering process can be approximated as a series of discrete complex reflectance coefficients spaced every 0.5 m in the fiber (the round-trip distance covered in 5 ns). Using this approximation, we can express the backscattered light at time $m$ as the sum of each input vector element scaled by the appropriate reflection coefficient: $\tilde{C}_m = \sum_{n=1}^{N} A_n \tilde{r}_{m-n+1}$, where $\tilde{r}_k = 0$ for $k < 1$ or $k > M$. In other words, the Rayleigh backscattering process performs a series of vector convolution operations, applying different random kernels to the input vector as the pulse train propagates down the fiber. We can also express this transform as a matrix vector multiplication: $\tilde{C}_M = \tilde{B}_{M \times N} \times A_N$, where $\tilde{B}_{M \times N}$ is an $M \times N$ transfer matrix. As shown in Fig. 2(c), the matrix $\tilde{B}$ contains the reflection coefficients, $\tilde{r}_k$, arranged so that each row in $\tilde{B}$ contains the same elements as the previous row, shifted by one column.

This system becomes particularly interesting if the fiber is longer than the equivalent length of the encoded pulse train (setting $M > N$). In this case the fiber can perform multiple, distinct random kernel operations on the input vector, effectively computing a grouped convolution on the same set of input data. This is significant from a neural network perspective since these grouped convolution operations are known to help extract hierarchical features in a dataset [16,17]. This is mathematically very different from most operations that have been implemented photonically in the past. First, photonic computing platforms exploiting random transforms [9,24,29] performed global, fully-random projections of the input data, rather than convolutions. Second, most photonic computing platforms, including highly reconfigurable integrated photonic systems [10,11], are limited to applying a single kernel in each layer, rather than performing multiple, distinct kernel transforms on the same input data. A notable exception is the time-wavelength multiplexed approach which leveraged frequency combs to perform grouped convolutions [18]. However, in addition to the complexity of this approach, the number of kernels was limited by the number of individually addressable comb teeth.

In the distributed feedback system, the number of kernel operations can be increased simply by using a longer optical fiber or a faster data encoding rate. In particular, the length of the output vector $\tilde{C}_M$ is set by the length of the fiber, $L$, and the data encoding rate, $f_0$, as $M \approx 2 \cdot ([2L/(c/n)]/\tau + N)$, where $c$ is the speed of light, $n$ is the effective index in the fiber, and $\tau = 1/f_0$ is the length of the pulses representing the elements of $A_N$. The term in the square brackets represents the round-trip time in the fiber while the factor of 2 outside the brackets accounts for using a polarization diversity receiver to record the backscattered light in both polarizations in parallel (not shown in Fig. 2, but used in the experiments described below, see Methods). The additional $N$ accounts for the length of the input vector and assumes we make use of RBS light that does not include every element in $A_N$ (i.e. RBS light collected before the entire pulse train enters the fiber and after the pulse train starts to leave the fiber). This expression also assumes that backscattered light is sampled at the data encoding rate of $f_0$ and that the temporal correlation width of the Rayleigh pattern matches the pulse duration $\tau$, which is the case for Rayleigh scattering [28]. By increasing the data encoding rate and the fiber length, this technique could be used to process large scale matrices or perform multiple distinct kernel operations on the same input vector (the number of distinct kernels is set by the ratio $M/N$). For example, this platform could perform MVMs with $M = 10^6$ using an encoding rate of 10 GHz and a 5 km fiber before attenuation becomes significant (note that 0.2 dB/km is typical for telecom fiber at a wavelength of 1550 nm, resulting in a round-trip loss of 2 dB for a 5 km fiber).

Although the temporal multiplexing approach presented here trades-off computing speed for the ability to use a single modulator and detector, the availability of high-speed optical modulators and detectors (e.g. 20

GHz devices are widely available) helps to mitigate this trade-off. The time required to compute a MVM can be expressed as $\tau_{MVM} = N/f_0 + 2L/(c/n)$, where the first term accounts for the length of the input pulse train and the second term represents the round-trip time in the fiber. In this work, we used a 500 m fiber and a 200 MHz encoding rate, yielding $\tau_{MVM} \approx 10\ \mu s$ for $N$ approaching $10^3$. Increasing the encoding rate to 20 GHz and using a 5 m fiber could enable a 100x speed-up ($\tau_{MVM} \approx 100\ ns$) while performing a MVM with the same matrix dimensions. This analysis implies that using higher frequency encoding is generally beneficial, enabling faster computation for a given matrix size. However, as we will discuss below, the power consumption also increases with the encoding rate and the proper balance will depend on the application.

**Non-linear Principal Component Analysis**

As an initial demonstration of this technique, we considered a text-book example of non-linear principal component analysis. Non-linear PCA is an unsupervised learning technique that has been used for dimensionality reduction, singular value decomposition, denoising, and regression analysis [30,31]. A standard PCA relies on linear transforms to project data onto a new coordinate system that represents the variance in a dataset using as few dimensions as possible. However, relying entirely on linear transforms limits a standard PCA to analyzing data that is linearly separable [32]. A non-linear PCA operates by first applying a non-linear transform to a dataset before performing a standard PCA, facilitating the analysis of a wider range of data types. To illustrate how our platform can be used for non-linear PCA, we first created a dataset consisting of 500 points (defined by their cartesian coordinates $x, y, z$) randomly distributed on 3 concentric spheres with radii of 1, 2, or 3, as shown in Fig. 3(a). Ideally, the PCA would decompose the output vector into a single non-zero principal component representing the length of the $x, y, z$ vector. At minimum, the PCA should result in <3 significant PCs representing a low dimensional space in which the data points can be linearly separated. A standard linear PCA is unable to separate these 3 classes of points, as shown in Fig. 3(b) which plots the weights of the first 2 PCs for each data point. Moreover, simply expanding the dimensionality without applying a non-linear transform was unable to separate the classes on its own. To illustrate this, we computationally applied a linear random transform to each coordinate by multiplying each coordinate by a 3x200 random matrix before applying a PCA. As shown in Fig. 3(c), this cascaded transform (linear dimension expansion followed by PCA) is still linear and is unable to separate the classes.

To perform a non-linear PCA, we used the Rayleigh backscattering platform to create a non-linear projection of the data onto a higher-dimensional space before applying a PCA. To do this, we injected each data point into the fiber using a train of three 50 ns pulses (i.e. an input vector with $N = 3$ and $\tau = 50\ ns$). The EOM was initially biased at zero transmission and the voltage sent to the EOM was set by the amplitude of the $x, y, z$ coordinates of each point (normalized such that the maximum coordinate of "3" was set to the maximum transmission voltage for the EOM, $V_\pi$). We then recorded the Rayleigh backscattered speckle pattern produced by each point, yielding a $M = 200$ output vector. This process effectively projected each 3-dimensional point into a 200-dimensional speckle pattern.

We then performed a standard linear PCA on the backscattered speckle patterns. As shown in Fig. 3(d), the hybrid photonic/electronic non-linear PCA efficiently separated the three classes of points. This separation relied on both expanding the dimensionality and applying a non-linear transform to the original data. To illustrate this, we also attempted to perform a PCA using just 3 of the speckle grains in each backscattered speckle pattern. As shown in Fig. 3(e), this was unable to efficiently separate the three classes (the separation depends on the 3 speckle grains selected, but the result shown is typical for most sets of 3 speckle grains). Expanding the dimensionality increased the chance of randomly finding a transform that projected the data into a space where it is highly separable.

While this demonstration showed that non-linear random projections can help identify variations in a dataset, the benefit of using analog optics (in terms of power consumption and speed) was limited since the final PCA was performed on a relatively large dataset with 200 dimensions. Fortunately, after expanding the dataset with the non-linear transform, we could then compress it into a lower dimensional vector in the analog optical domain before performing the final PCA while maintaining many of the benefits. To test this, we re-compressed the 200-dimensional RBS speckle pattern produced by each point to 3-dimensions by averaging 100 points at a time. Experimentally, this could be achieved using a photodetector with a low-pass filtering response. As shown in Fig. 3(f), a PCA performed on the re-compressed vectors was still able to efficiently separate the three classes. This simple demonstration illustrates how a non-linear random projection can be used to identify key features in a dataset.

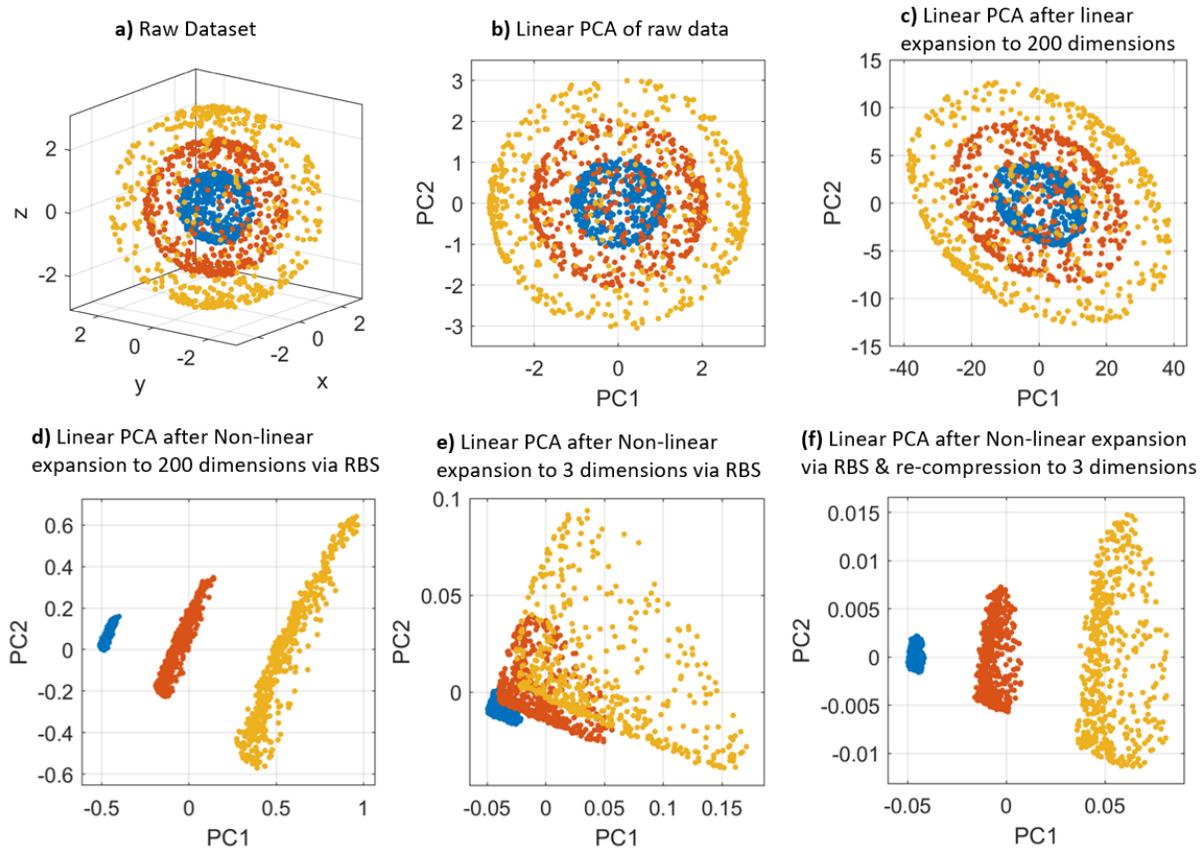

**Figure. 3 Non-linear PCA using distributed feedback.** (a) The raw dataset consisted of 500 3-dimensional points randomly distributed on 3 concentric spheres. (b) Linear PCA of the raw data fails to separate the 3 classes of points. Plotted are the weights of the first 2 principle components for each datapoint. (c) A random linear transform is similarly unable to separate the 3 classes of points. (d) After the application of a non-linear random projection using Rayleigh backscattering, the 3 classes are clearly separable using a standard PCA. (e) Result of a PCA applied to 3 speckle grains selected from the RBS pattern, showing that the classes are difficult to separate without expanding the dimensionality. (f) Result of a PCA applied after re-compressing the RBS speckle pattern into 3 dimensions using a low-pass filter.

**Non-linear Support Vector Machine**

The same non-linear random projections can be used to construct a non-linear support vector machine (NL-SVM). SVMs are a supervised learning technique designed to separate different classes in a dataset by identifying the maximum margin hyperplane separating two classes. A non-linear SVM first projects the data into a higher-dimensional space before finding a hyperplane to separate different classes [33,34]. The platform proposed in this work is ideally suited for non-linear SVM, since the grouped convolutions introduced by the RBS process can efficiently project an input dataset into a higher-dimensional space to facilitate classification.

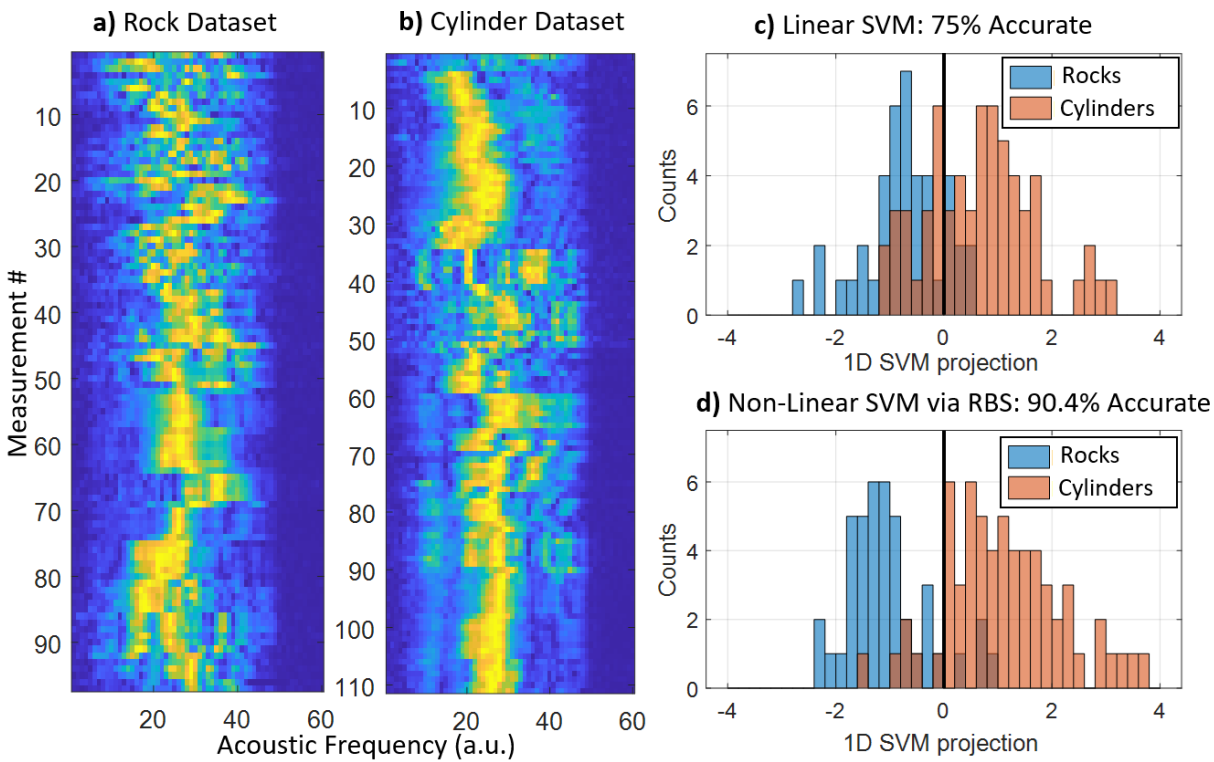

**Figure 4. Non-linear SVM using distributed feedback.** The SONAR dataset consisted of 97 measurements of rocks (a) and 111 measurements of cylinders (b). (c) Applying a linear SVM directly to the training data resulted in a classification accuracy of 75%. (d) After using the optical platform to perform a non-linear random projection on the SONAR data, the SVM accuracy increased to 90.4%.

To explore the use of our platform for NL-SVM, we selected a benchmark SONAR dataset consisting of SONAR measurements of either rocks or metal cylinders (phantoms for underwater mines) [35]. The dataset consists of 97 measurements of rocks and 111 measurements of cylinders. Each measurement was obtained using a frequency-modulated chirped SONAR and contains a 60-element vector representing the reflected signal as a function of acoustic frequency. This dataset has been divided into a training and a validation dataset, each containing 104 measurements (we focused on the "Aspect Angle Dependent" test described in [35]). We first attempted to classify this data by using a standard linear SVM to assign a hyperplane based on the training data and then evaluated how well this hyperplane separates the validation data. A histogram showing the position of the validation data measurements along a 1-dimensional SVM projection is shown in Fig. 4(c). Ideally, the two classes would be completely separable in this space; however, we observed significant overlap between the classes and obtained an overall classification accuracy of only 75%. We then used the RBS platform to perform a random non-linear projection on the SONAR data before applying an SVM. In this case, each SONAR measurement (consisting of a 60-element vector) was encoded in a pulse train using 5 ns pulses. An example of an encoded pulse train and the resulting RBS pattern are shown in Fig. 2(d,e) The RBS process was used to expand the dimensionality of the SONAR signal from 60 to 2000. As shown in Fig. 4(d), the non-linear SVM was much more effective at separating the two classes, obtaining an accuracy of 90.4% on the validation data, comparable to the performance of the neural network reported in [35]. This illustrates the potential for our platform to facilitate data classification by transforming input data into higher dimensional space using grouped random convolutions.

**Extreme Learning Machine**

Extreme learning machines are a type of feed-forward neural network in which the weights and connections in the hidden layers are fixed and a single decision layer is trained to complete a task [36,37]. ELMs were initially proposed to avoid the computational demands of training every connection in a neural network, but their unique structure is particularly well-suited for optical implementations. Photonic ELMs can use passive photonic structures to apply a complex transform (i.e. the fixed layer in the ELM) and rely on a single electronic decision layer to complete the computing task [23,38]. This has enabled photonic computing architectures built around multimode fiber [23] or complex disordered materials [19] where precise control of the transfer matrix would be challenging. Here, we show that our distributed optical feedback platform can be configured as an ELM to perform image classification.

We tested our system on two benchmark tasks: classifying the MNIST Digit database and the MNIST fashion database [39]. Each dataset consists of 60,000 training images and 10,000 test images in 10 classes (either hand-written digits from 0-9 or 10 types of clothing). Although the distributed feedback system performs vector convolutions, it can also be used to perform two-dimensional convolutions on image data that has been flattened to a one-dimensional vector (see Methods for details) [18]. To do this, we encoded each image as a one-dimensional pulse train by assigning the magnitude of each 5 ns pulse to the intensity of one pixel in the image. We injected pulse trains representing all 70,000 images in series and recorded 70,000 Rayleigh backscattered speckle patterns. We then used a ridge regression algorithm to train a decision layer to classify the 60,000 training images. Finally, we tested the ELM using the backscattered patterns obtained from the 10,000 validation images. This same process was repeated for the digit and fashion databases.

As shown in Fig. 5(a,b), the distributed feedback ELM system achieved an accuracy of 96.7% and 85.3% on the digit and fashion databases, respectively. This performance is comparable to the best performing photonic neural networks [38] (but at higher speed than competing free-space architectures) and to digital electronic neural networks with similar depth [39] (but with reduced power consumption, as detailed in the next section). Our approach is able to achieve this accuracy without training the underlying transforms by expanding the dimensionality and leveraging the inherent power of grouped convolutions to extract high-level features from a dataset. This demonstration also shows that the same platform, and, in fact, the same optical fiber, can be used for multiple tasks simply by re-training the decision layer, consistent with previous demonstrations of the versatility of photonic ELMs [23,38].

The most important parameter impacting the ELM accuracy is the length of the Rayleigh backscattering pattern (i.e. the length of the output vector, or the extent to which the system expanded the dimensionality of the original data). The entire measured RBS pattern consisted of ~2500 speckle grains (a ~6x increase in dimensionality compared to the 400-pixel MNIST Digit images). To investigate this dependence, we sub-sampled the measured RBS pattern and repeated the training and test procedure using output vectors of varying length. As shown in Fig. 5(c), the accuracy increases rapidly with the length of the output vector before gradually plateauing. Higher accuracy is achieved for the digit database, which is consistent with previous studies showing that the fashion database is more challenging [39]. Nonetheless, the accuracy increases monotonically with the size of the output dimension in both cases. Fortunately, the distributed feedback platform is well suited for this task and can increase the dimensionality of the output data simply by using a fiber that is longer than the effective length of the pulse train. Moreover, since the backscattering process is entirely passive, increasing the dimensionality in this way has a negligible effect on the power consumption despite increasing the size of the MVM.

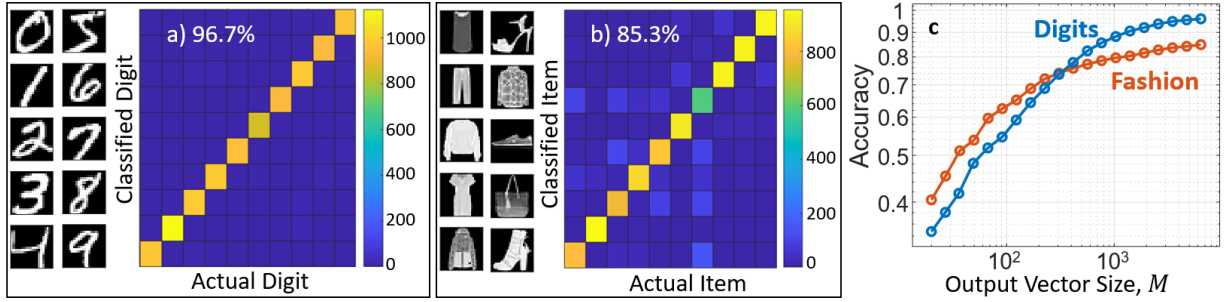

**Figure 5. Extreme learning machine using distributed feedback.** (a,b) The photonic ELM classified the MNIST digit database with an accuracy of 96.7% and the Fashion database with an accuracy of 85.3%. (c) The ELM accuracy increases with the dimensionality of the output vector, $M$.

**Energy Consumption**

As discussed, the distributed feedback process effectively performs a grouped convolution on the input vector, which can be expressed as a matrix vector multiplication. This framework allows us to compare the power consumption using our approach to a standard digital electronic processor like a GPU by analyzing the energy per multiply-accumulate (MAC) operation. Our analysis accounts for the energy required to operate the laser, the modulator, and the photodetector. Assuming shot-noise limited detection, we first estimated the optical power required at the detector to obtain a Rayleigh backscattering pattern with the desired signal-to-noise ratio (SNR), as [40]:

$$P_{Rx} = 2^{2enob} q f_0 / \mathcal{R} \quad [1]$$

where $enob$ is the effective number of bits ($SNR = 6.02 \times enob + 1.76$ in dB), $q$ is the charge of an electron, $f_0$ is the measurement bandwidth (and the data encoding rate), and $\mathcal{R}$ is the responsivity of the detector. We can then estimate the required laser power as

$$P_{laser} = P_{Rx} / [T_{mod} r_{RBS}] \quad [2]$$

where $T_{mod}$ is the transmission through the modulator and $r_{RBS}$ is the average reflectance due to Rayleigh backscattering in the fiber. Since Rayleigh backscattering is a distributed process, the effective Rayleigh backscattering coefficient depends on the duration of the pulse train launched into the fiber and can be estimated as

$$r_{RBS} = (-82 dB/ns)(N \cdot \tau) \quad [3]$$

for Corning SMF-28e+ (as used in this work) [41], where $N$ is the length of the input vector and $\tau$ is the pulse duration for each element in the input vector. The total electrical power required to operate the laser can then be calculated as $P_{laser}/\eta$, where $\eta$ is the wall-plug efficiency of the laser.

The electrical power consumed by the modulator can be expressed as [42]

$$P_{mod} = \frac{N}{M}\left(\frac{1}{2} C_{mod} V_\pi^2 f_0\right) \quad [4]$$

where $N$ is the length of the input vector, $M$ is the length of the output vector (i.e. the Rayleigh backscattering pattern), $C_{mod}$ is the capacitance of the modulator, and $V_\pi$ is the peak-to-peak driving voltage of the modulator. The ratio of $N/M$ denotes the operating duty-cycle of the modulator and we assume $M \geq N$, which corresponds to expanding the dimensionality of the input vector. The electrical power consumed by the photodetector can be expressed as

$$P_{det} \approx V_{bias} \mathcal{R} P_{Rx} \qquad [5]$$

where $V_{bias}$ is the bias voltage applied to the detector. The total power can then be expressed as $P_{total} = P_{laser}/\eta + P_{mod} + P_{det}$. In order to estimate the energy per MAC, we can then calculate the total number of MACs per second as $N \cdot M / \tau_{MVM}$, where $\tau_{MVM}$ is the time required to complete a matrix vector multiplication. Since a MVM is completed in the time required to record a Rayleigh backscattering pattern with $M$ outputs, $\tau_{MVM} = M \cdot \tau$. Thus, the number of MACs per second is $N/\tau$ or $Nf_0$ and the energy per MAC can be estimated as $P_{total}/(Nf_0)$. Based on this analysis, we find that the energy per MAC required to power the laser scales as $f_0/N^2$ whereas the energy per MAC required to power the modulator and detector scales as $1/N$ and is independent of modulation frequency. This difference in scaling results from the dependence of $r_{RBS}$ on $N$ and $\tau$ in Eq. 3. Finally, this analysis ignored the use of a polarization diversity receiver for simplicity. However, using polarization diversity does not change the energy per MAC. For example, using polarization diversity would require twice the laser power to achieve the same SNR, since the RBS light is split between two detectors, but would enables twice the MACs per second. Instead, using a polarization diversity receiver doubles the data throughput (for the same modulation and detection speeds) without affecting the power efficiency.

This model allowed us to quantitatively evaluate the energy/MAC which could be achieved using the optical distributed feedback scheme. Here, we assumed a required $enob = 6$, which corresponds to a SNR of 38 dB, considerably higher than the 23.6 dB SNR of the RBS patterns measured experimentally in this work (see Methods). We set $M = 10 \times N$, implying that the system was used to expand the dimensionality of the input vector by a factor of 10. We then assumed typical values for the remaining parameters: a laser with wall-plug efficiency of $\eta = 0.2$, a modulator with $C_{mod} = 1\,fF$ and $V_\pi = 1\,V$, and a detector with $V_{bias} = 3V$ and $\mathcal{R} = 1\,A/W$ [43,44]. We first estimated the energy per MAC as a function of the input vector size $N$ using a modulation frequency of 200 MHz, matching our experimental conditions. As shown in Fig. 6(a), the energy/MAC is dominated by the laser under these conditions. For comparison, Fig. 6 also shows the power consumption typical of state-of-the-art GPUs (~1 pJ/MAC), which is independent of vector size [25]. At a vector size of $N = 784$, corresponding to the size of the $28 \times 28$ pixel images in the fashion database, the RBS system provides a $30 \times$ reduction in power consumption compared to a GPU. This improvement increases dramatically for larger vector sizes due to the $1/N^2$ dependency of the laser energy. Reducing the required $enob$ would also significantly reduce the power consumption by reducing the required laser power.

For many applications, using higher modulation frequencies could be attractive to enable low latency computing and higher data throughput. The expected energy/MAC at an encoding rate of 10 GHz is shown in orange in Fig. 6(b) (the laser is still the dominant source of energy consumption). In this system, the energy/MAC increases with encoding rate to compensate for the reduced backscattering at higher encoding rates (since $r_{RBS}$ depends on $\tau$). Nonetheless, the RBS system still outperforms a GPU for input vectors with $N > 10^3$ at an encoding rate of 10 GHz. In order to reduce the energy consumption, we could use enhanced backscattering fiber, which is commercially available and provides ~15 dB higher backscattering than standard single-mode fiber [45]. Figure 6(b) also shows the energy/MAC required using enhanced backscattering fiber with a data encoding rate of 10 GHz, which approaches the energy/MAC achieved at

200 MHz using standard fiber. Combining enhanced scattering fiber with a 200 MHz encoding rate could enable even lower power consumption, outperforming a GPU for vectors with $N > 25$. Although this type of fiber was initially developed for fiber optic sensing, it is ideally suited for this type of optical computing platform and could enable a significant reduction in power consumption. A distributed feedback fiber employing ultra-weak fiber Bragg gratings [46] or point reflectors [47] could also be used to achieve higher reflectance and thus lower power consumption. These technologies could also enable customized kernel transforms by tailoring the position and strength of each reflector.

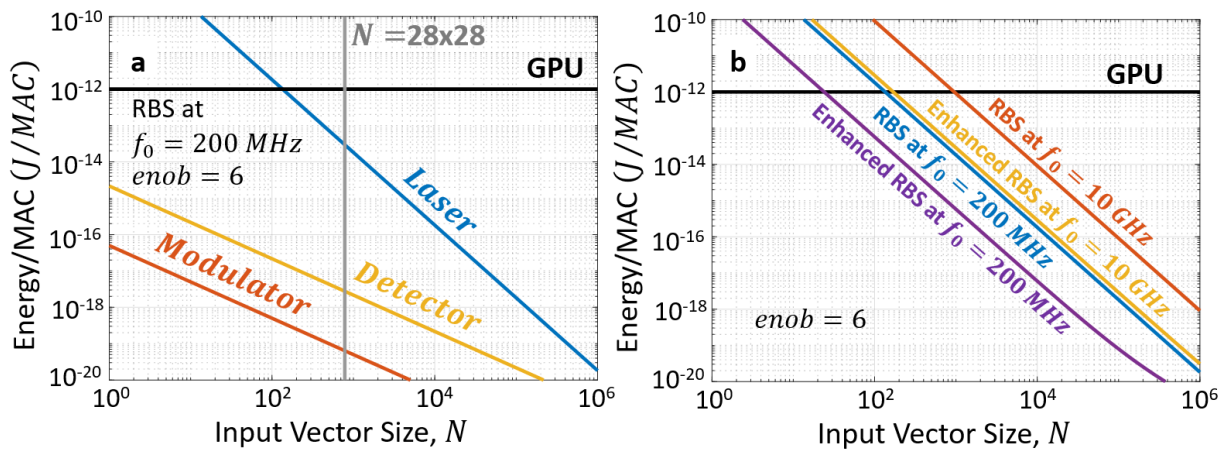

**Figure 6. Power Consumption.** (a) The energy/MAC required by the laser, detector, and modulator as a function of input vector size at an encoding rate of 200 MHz and assuming an $enob$ of 6. (b) The total energy/MAC for varying encoding rates, $f_0$, using either standard fiber (i.e. relying on Rayleigh backscattering, RBS) or using enhanced Rayleigh backscattering fiber with 15 dB higher reflectance.

**Discussion**

One challenge not addressed in this initial work is the environmental stability of the fiber. The PCA, SVM, and ELM algorithms investigated here each assume that a given launch pattern will always produce the same Rayleigh backscattering pattern (albeit with a finite SNR). However, the Rayleigh backscattering pattern depends on temperature and strain in addition to the launch pattern (Rayleigh backscattering in optical fiber is often used for distributed temperature and strain sensing [48]). As an example, for the $N = 400$ pixel images in the digit database encoded at 200 MHz, the Rayleigh backscattering pattern would decorrelate if the temperature of the fiber drifted by 0.4 $m°C$ (see Methods for details). While this level of temperature stabilization is possible, single-mode fiber is a very controlled platform with limited degrees of freedom and relatively simple calibration techniques have been proposed in other applications leveraging Rayleigh backscattering which might be applicable here [49]. Note that in the experiments reported in this work, the fiber was enclosed in an aluminum box, but was not temperature stabilized. Instead, the training and validation datasets were recorded during a short measurement window while the fiber environment remained stable. For example, the ELM experiments required 0.7 seconds to record the RBS pattern from all 70,000 test images and we found that the RBS pattern did not drift significantly on this time-scale.

While this work introduced a basic architecture for performing optical computing with distributed feedback, there are several areas for improvement and further investigation. First, introducing non-linearities in the optical domain could enable more sophisticated, multi-layer networks capable of addressing more challenging computing tasks [50]. There are a few promising approaches which could be pursued to introduce intra-network non-linearities in this platform. For example, an analog optical-electrical-optical conversion could be introduced, following the approach used in single-node fiber-optic reservoir computers [15] and recently explored in the context of integrated photonic neural networks [50,51]. Alternately, the long interaction lengths and high-power density in single mode fiber could be leveraged to introduce all-optical non-linearities such as such as four-wave mixing, modulation instability, or stimulated Brillouin scattering [52]. Second, the temporal multiplexing scheme proposed here sacrifices computing speed for simplicity by enabling the use of a single modulator to encode the entire input vector. In the future, some degree of spatial multiplexing could also be included to balance this trade-off. On the detection side, this is particularly straight-forward and was already implemented in part using a polarization diversity receiver. In the future, recording the backscattered pattern from a few-mode fiber or multiple discrete fibers could increase the output vector size without compromising the computing speed. Some degree of parallelization could also be explored on the data encoding side, e.g., by using multiple modulators and coupling light into a few-mode fiber. Third, while we focused on random projections in this work, the same basic combination of temporal encoding and distributed feedback could be used to perform other operations. For example, instead of relying on Rayleigh backscattering, a series of carefully positioned partial reflectors (such as weak fiber Bragg gratings [46] or point reflectors [47]) could be used to implement distinct kernel transforms at different positions along the fiber (inverse design principles could be used to optimize the reflector geometry).

In summary, this work introduced an optical computing platform based on temporal multiplexing and distributed feedback that performs grouped convolutions using a passive optical fiber. We showed that Rayleigh backscattering in single mode fiber can be used to perform non-linear random kernel transforms on arbitrary input data to facilitate a variety of computing tasks, including non-linear principal component analysis, support vector machines, or extreme learning machines. This approach enables large scale MVMs with $O(N)$ energy scaling using a single modulator and photodetector. The entire system can be constructed using off-the-shelf fiber-coupled components, providing a compact and accessible approach to analog optical computing. Finally, since this approach operates on temporally encoded light in standard single-mode fiber, it could potentially be applied directly to optical data transmitted over fiber enabling applications in remote sensing and RF photonics.

**Methods**

Experimental Details: In the experiments reported in this work, an Erbium-doped fiber amplifier (EDFA) was inserted after the EOM in Figure 2 to increase the launch power and compensate for the use of a relatively low power (10 mW) seed laser and relatively high insertion loss EOM (4.5 dB). In the future, a higher power laser could be used to avoid needing the EDFA (depending on the length of the pulse train and encoding rate, peak launch power in the range of 10-100 mW is sufficient to achieve an *enob* of 6 based on equations 1-3). In this work, we also used a second EDFA followed by a wavelength division multiplexing filter after the circulator to amplify the Rayleigh backscattered pattern before detection. This EDFA was used to minimize the effect of photodetector noise. In the future, using a lower-noise detector would preclude the need for the second EDFA. Note that the slow response time of the EDFA (~ms) prevented it from introducing a non-linear response on the time-scale of the RBS pattern.

In the ELM experiments, the 2-dimensional images were flattened into 1-dimensional vectors. As discussed in [18], vector convolutions can be used to perform a convolution on image data, although there is an "overhead" cost (not all output measurements are used) and the "stride" is inherently asymmetric (a symmetric stride could be obtained through sequential measurements by modifying the encoding strategy). In the image recognition tasks described here, we used all of the output samples, which included measurements representing standard 2D convolutions as well as mixtures that convolved different combinations of input pixels and would not be computed in a standard convolution. We also did not attempt to correct for the asymmetric stride introduced by the vector convolution process.

The digitized RBS speckle patterns were recorded at 1 GS/s. This is slightly faster than the temporal correlation width, which is set by the data encoding rate (i.e. 200 MHz for the SVM or ELM test and 20 MHz for the PCA test) [28]. In this work, we processed slightly oversampled Rayleigh backscattering patterns, but the dimensionality listed as $M$ in each section corresponded to the pattern length divided by the correlation width of 5 ns or 50 ns.

We calculated the SNR of the RBS patterns recorded in this work as the ratio of the average power of the RBS pattern to the standard deviation in the background (recorded with the EOM blocked). We found an SNR of 23.6 dB, which corresponds to an *enob* of 3.6. Fortunately, neural networks are quite robust to low-precision computing [53], enabling the excellent performance achieved in the benchmark tasks explored in this work despite this modest precision.

Temperature Sensitivity: The temperature sensitivity of the Rayleigh backscattering pattern depends on the length of the incident pulse train and can be estimated as $(f_0/N)/[1.2 GHz/°C]$, where the $(f_0/N)$ term represents the transform limit of the input pulse train (which has a duration of $N\tau = N/f_0$) and the term in square brackets represents the shift in the Rayleigh spectrum with temperature [54]. This expression calculates the temperature shift required for the Rayleigh backscattering spectrum to shift by the transform limit of the pulse train, which would result in a decorrelated speckle pattern.

**Acknowledgement**

R.S. acknowledges the support by the Laboratory Directed Research and Development program at Sandia National Laboratories, a multimission laboratory managed and operated by National Technology and Engineering Solutions of Sandia, LLC, a wholly owned subsidiary of Honeywell International, Inc., for the U.S. Department of Energy's National Nuclear Security Administration under contract DE-NA-003525.This paper describes objective technical results and analysis. Any subjective views or opinions that might be expressed in the paper do not necessarily represent the views of the U.S. Department of Energy or the United States Government.